\documentclass[11pt]{article}
\usepackage{geometry}
\usepackage[utf8]{inputenc}
\usepackage{graphicx}
\usepackage{amsmath}
\usepackage{amssymb}
\usepackage{authblk}
\usepackage[english]{babel}
\title{Creation of NV centers in diamond under 155 MeV electron irradiation}

\author[1,2]{Elena Losero}
\author[2]{Valentin Goblot}
\author[2]{Yuchun Zhu}
\author[2]{Hossein Babashah}
\author[3]{Victor Boureau}
\author[4]{Florian Burkart}
\author[2]{Christophe Galland}

\affil[1]{Division of Quantum Metrology and Nanotechnologies, Istituto Nazionale di Ricerca Metrologica (INRiM), Strada delle Cacce 91, 10135 Torino, Italy}
\affil[2]{School of Basic Sciences, EPFL, Rte Cantonale, 1015, Lausanne, Switzerland}
\affil[3]{Interdisciplinary Center for Electron Microscopy (CIME), EPFL, Rte Cantonale, 1015, Lausanne, Switzerland}
\affil[4]{Deutsches Elektronen-Synchrotron DESY, Hamburg, Germany}

\begin{document}

\maketitle

\section*{Abstract}
Single-crystal diamond substrates presenting a high concentration of negatively charged nitrogen-vacancy centers (NV$^-$) are on high demand for the development of optically pumped solid-state sensors such as magnetometers, thermometers or electrometers. While nitrogen impurities can be easily incorporated during crystal growth, the creation of vacancies requires further treatment. 
Electron irradiation and annealing is often chosen in this context, offering advantages with respect to irradiation by heavier particles that negatively affect the crystal lattice structure and consequently the NV$^-$ optical and spin properties. A thorough investigation of electron irradiation possibilities is needed to optimize the process and improve the sensitivity of NV-based sensors. 
In this work we examine the effect of electron irradiation in a previously unexplored regime: extremely high energy electrons, at 155~MeV. We develop 
a simulation model to estimate the concentration of created vacancies and experimentally demonstrate an increase of NV$^-$ concentration by more than 3 orders of magnitude following irradiation of a nitrogen-rich HPHT diamond over a very large sample volume, which translates into an important gain in sensitivity. Moreover, we discuss the impact of electron irradiation in this peculiar regime on other figures of merits relevant for NV sensing, i.e. charge state conversion efficiency and spin relaxation time. Finally, the effect of extremely high energy irradiation is compared with the more conventional low energy irradiation process, employing 200~keV electrons from a transmission electron microscope, for different substrates and irradiation fluences, evidencing sixty-fold higher yield of vacancy creation per electron at 155~MeV.

\section{Introduction}
Negatively charged nitrogen vacancy (NV$^-$) centers in diamond have raised a lot of attention in the last decades for their promising sensing capabilities under ambient conditions, particularly for measuring magnetic fields, electric fields and temperature \cite{rondin2014magnetometry,jensen2017magnetometry,radtke2019nanoscale,rembold2020introduction}. The sensing mechanism is typically based on optical polarization and readout of their the spin states \cite{schirhagl2014nitrogen}.
Both single NVs \cite{degen2008scanning,balasubramanian2008nanoscale,maze2008nanoscale} and ensemble \cite{taylor2008high,acosta2009diamonds} can be used. The first approach offers atomic spatial resolution and high readout contrast, but the photon flux is low and limits the absolute sensitivity. On the other hand, NV ensembles are an attractive option for large scale applications: in this case the spatial resolution is given by the optically probed volume and the extracted signal scales with $N$, while the photon shot noise with $\sqrt{N}$,  $N$ being the number of excited NV centers. Therefore, having high NV density is of utmost importance for achieving high signal to noise ratio and sensitivity \cite{wolf2015subpicotesla}. 

A lot of efforts have been invested in engineering diamond with high NV concentration and long coherence time for quantum sensing applications. One promising approach focuses on tailoring NV properties during the growth of diamond itself \cite{achard2020chemical,tallaire2020high,jaffe2020novel}.
However, despite impressive progress in this direction, these processes are still not standard. They are typically employed for thin layers only and result in expensive end products.
Another possible approach consists in increasing the NV density inside more affordable commercially available diamond substrates, synthesized with conventional techniques. This is the approach we focus on in this work.

In order to create new NV centers in a pristine diamond substrate, substitutional nitrogen atoms (N) have to bond with vacancies (V) in the diamond lattice. Nitrogen is naturally present in most commercially available diamond samples, with very different concentrations [N] depending on the growth technique \cite{ashfold2020nitrogen}. Note that throughout this work, we mainly use ppm as concentration measurement unit. Considering diamond lattice parameters, we recall that 1~ppm = $1.75 \cdot 10^{17}$ atoms$\cdot$cm$^{-3}$. 
High-pressure-high-temperature (HPHT) diamonds (type Ib) are low cost and high yield and typically present a large amount of non-intentionally doped nitrogen ([N]$\sim$10–300 ppm), while chemical vapour deposition (CVD) synthesis allows for a better control of nitrogen concentration. 'Optical grade' CVD-grown  substrates typically present nitrogen concentrations of few ppm and are routinely produced by many suppliers.

In contrast, all commercially available diamond substrates usually contain negligible amount of vacancies. The most common strategies for vacancy creation are ion irradiation \cite{haque2017overview,ohno2014three,huang2013diamond,acosta2009diamonds}(C, N, He, H, etc.) and electron irradiation \cite{beveratos2001nonclassical,boudou2009high,acosta2009diamonds,mclellan2016patterned,dantelle2010efficient,luo2022creation,wang2023enhancing}. Additional techniques were proposed and experimentally demonstrated, including laser writing \cite{chen2017laser} and neutron irradiation \cite{mita1996change}, among others.
Following irradiation, an annealing process allows vacancies to move and bond with the substitutional nitrogen atoms, converting a fraction of each of these two constituents into desired NV centers. Vacancies migration is typically achieved through thermal annealing above 600$^\circ$C, which is the threshold temperature at which they become mobile \cite{davies1992vacancy}. Standard annealing process consists in few hours at 700-900$^\circ$C. Low pressure is used to avoid graphitization. Annealing may also be performed directly during irradiation, to lower the formation of vacancies complexes \cite{capelli2019increased,deslandes2015diamond,mindarava2020efficient,choi2017depolarization}. 

Vacancy creation is thus the main tool for engineering NV centers post diamond synthesis. In particular, electron irradiation offers advantages compared to heavier particles in terms of the induced lattice damage \cite{barry2020sensitivity}.
Historically, electron irradiation was first performed at acceleration energy of few MeV (2-10~MeV), with large beams uniformly irradiating all the sample area and resulting in penetration depth up to a few millimeters. Later, low energy irradiation was explored using transmission electron microscopes (TEM), with electrons typically accelerated at 200~keV \cite{kim2012electron,farfurnik2017enhanced,mclellan2016patterned,schwartz2012effects}. This value is close to the minimum electron energy necessary to create a vacancy, estimated as $\sim$165~keV in \cite{campbell2000radiation}.
In this energy regime the penetration depth and the beam diameter are typically lower (few tens of $\mu$m), but these features can be used to create NV ensembles located in limited volumes. Moreover, TEM microscopes are available in many institutes, making the process widely accessible.
However, the regime of extremely high energy irradiation, above 100~MeV, remains unexplored. With the growth of applications for quantum sensing with NV ensembles, this regime could present interesting opportunities for NV center engineering, in particular for high density ensembles over macroscopic volumes.

\vspace{5mm}
Here, we investigate the effect of extremely high energy (155~MeV) electron irradiation of bulk diamond for NV creation and NV sensing application.
First, we develop a new simulation tool, able to evaluate the number of vacancies created by electrons with energy above 100~MeV. 
Simulation of vacancy creation is crucial to set the optimal irradiation parameters. As a rule of thumb, the concentration of created vacancies should be of the order of half the concentration of substitutional nitrogen present in the substrate, so that all vacancies can bond to a nitrogen atom and trap a free electron from a second nitrogen atom \cite{barry2020sensitivity}. Without sufficient nitrogen to act as electron donors, most NV are in their neutral charged state NV$^0$, which cannot be used in spin-based sensing protocols. 
In most experimental paper the vacancy concentration [V] is extrapolated from a couple of theoretical papers \cite{campbell2000radiation,campbell2002lattice}, where simulation was limited to the regime of few MeV. These results cannot be extrapolated to much higher electron energy, in particular due to the onset of secondary cascade processes. Our model describes vacancy creation in all energy regimes, from few keV to several hundreds of MeV.

We then perform electron irradiation with electrons at 155~MeV, on a commercially available HPHT sample ([N] $\lesssim$ 200 ppm). Irradiation is performed in a linear accelerator at the Deutsches Elektronen-Synchrotron (DESY). 
We characterize the obtained sample using different techniques to confirm that the lattice quality is not degraded and to quantify the sensing performances of the created NV centers. We report an increase in NV$^-$ density by more than a factor 2000, up to 0.6~ppm, over the entire sample length (3 mm) along the beam propagation axis, and unaffected optically detected magnetic resonance (ODMR) contrast and linewidth. This corresponds to an increase in projected sensitivity by a factor $\sim 45$ with respect to the sample before irradiation.

Finally, the results are also compared with those obtained using low energy irradiation. Two different substrates, HPHT and CVD, presenting  different levels of initial nitrogen concentration ([N]$\lesssim$200~ppm and [N]$\lesssim$1~ppm, respectively) are investigated. A TEM is used to irradiate several areas (diameter $\sim 15~\mu$m) on the substrates, with 200~keV electrons and different irradiation fluences. For a same substrate and fluence, we observe that the 155~MeV irradiation provides $\sim$60 times higher NV$^-$ concentration, in qualitative agreement with our simulation.
Moreover, we show in the 200~keV-irradiated samples that conversion efficiency from V to NV$^-$ saturates as electron fluence is increased, with a saturation threshold fixed by [N], and clearly identify the presence of the same mechanism in the 155~MeV-irradiated sample.

Overall, our work opens a new path of exploration for NV center engineering in diamond with electron irradiation, considering a new electron energy irradiation regime. A peculiar aspect of the ultra-high energy regime is that it allows to create NV$^-$ ensembles well suited to quantum sensing over macroscopic volumes: we estimate that vacancies can be efficiently generated through more than 10~cm of diamond, permitting massive saving in batch irradiation costs if hundreds of substrates are stacked together, for example.

\section{Simulation}\label{sec:Simulation}

\begin{figure}
    \centering
    \includegraphics[width=0.9\textwidth]{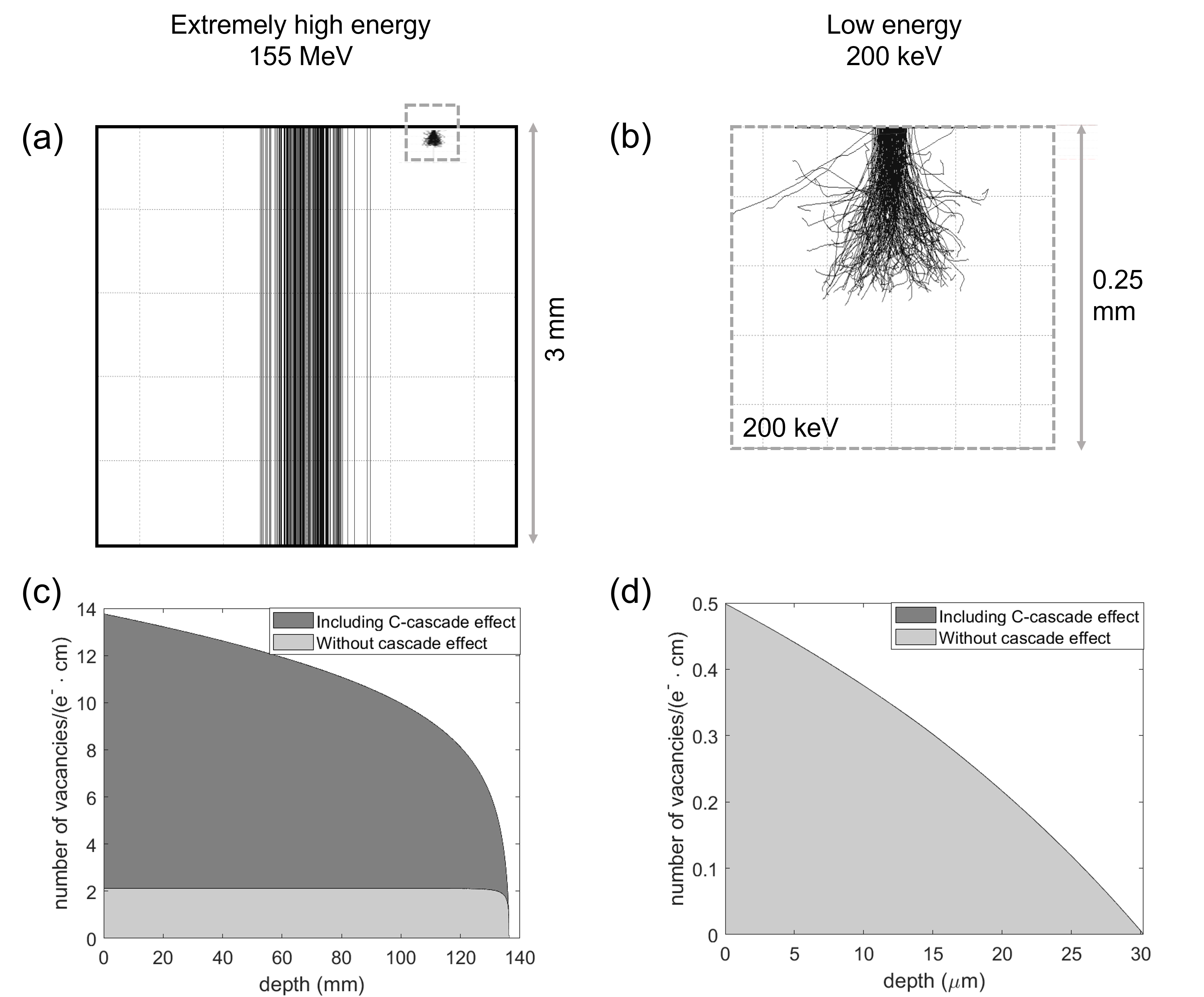}
    \caption{a-b) Trajectories of 200 electrons in diamond, simulated using CASINO software. a) 155 MeV electron irradiation, 500 $\mu$m beam diameter. b) 200 keV electron irradiation, 15 $\mu$m beam diameter. c-d) Simulation of the number of vacancies per centimeter depth generated by 1 accelerated electron at c) 155 MeV energy and d) 200 keV energy. The contribution of the initial recoiled atoms (filled light gray) and the secondary cascade process (filled dark gray) are highlighted. Note the different axes limits in the two cases.}\label{fig:Simulation}
\end{figure}

Electron trajectories inside diamond can be computed with the CASINO software, a freely available Monte Carlo simulation tool \cite{drouin2007casino}. We use it to estimate the electron penetration depth. The results obtained are reported in Fig. \ref{fig:Simulation} a and b, for extremely high energy (155~MeV) and low energy (200~keV) respectively. 200 electron trajectories are displayed in both cases. In the first case, almost all electrons go through the 3 mm diamond in a straight line, due to the very high electron energy.
In the second case, their low kinetic energy does not allow electrons to exit from the sample, the penetration depth being limited to slightly more than 100 $\mu$m. Vacancies (and thus NV centers) are here created in a shallower layer only, limited by the minimum energy that an electron should have to displace a carbon atom. 

However, CASINO does not model the generation of crystal defects induced by the electron beam interaction with the sample. In particular, the density of created vacancies cannot be directly estimated with this software. Especially at very high energies, the effect of the displaced carbon atoms is expected to play a crucial role since they have enough energy to displace other atoms through a cascade phenomenon \cite{campbell2000radiation,davies2001interstitials}. This secondary process is not implemented in CASINO, which only considers the primary electron-carbon interaction.
In \cite{campbell2000radiation} a different Monte Carlo simulation tool is proposed, to discuss the damage induced by electrons in diamond, in the range from 0.25 to 10 MeV. 
To explore the effect of electron irradiation in the regime of ultra high energy, where radiative energy losses of electrons and vacancies generated by cascade phenomenon are prevailing, we develop an in-house model that is able to provide reliable results in all energy regimes from few keV to hundreds of MeV. 

The simulation has been implemented using the proprietary coding language of Gatan Microscopy Suite (GMS 3.5) software and is available upon request. First, the energy profile of the electron beam is calculated versus the propagation depth in the material. For this purpose, the energy losses by ionization, predominant at low beam energies (e.g. 200~keV), and the radiative energy losses by bremsstrahlung, predominant at very high beam energies (e.g. 155~MeV), are considered. Second, the density of lattice vacancies is calculated versus the depth in the diamond crystal. For this purpose, the energy transfer events from the fast electrons to the carbon nuclei are considered. If the energy transfer is higher than the threshold energy capable of ejecting an atom from the crystal lattice, a vacancy-interstitial pair is created in the crystal. This displacement threshold energy is around 35~eV in diamond \cite{campbell2000radiation}, and corresponds to the maximal energy transferable by a 165~keV electron. 
If a recoiled carbon atom has an energy higher than the displacement threshold energy, it can then generate further vacancies through atom-atom collisions, known as cascade process. It has been predicted with molecular dynamics that 50\% of the displaced carbon atoms in cascade process recombine and do not generate stable vacancies in diamond \cite{buchan_molecular_2015}; this phenomenon is included in our simulation. The details of the model are described in the Supplementary Information.

Figs.~\ref{fig:Simulation}c and d show the number of vacancies per centimeter depth generated by a single accelerated electron. We observe that the vacancy generation through cascade process is not activated for 200~keV beam while it is predominant for 155 MeV beam. As complementary information, in Fig.~\ref{fig:energyprofiles}b the number of generated vacancies versus the beam energy is reported. The 200~keV electrons are not able to generate vacancies after a propagation of 38~$\mu$m as the maximal energy transferable to a carbon atom is decreased below the displacement threshold energy at this depth.
The vacancy generation of the 155~MeV beam is almost constant after a propagation depth of 3~mm in diamond, with a beam energy decreased to 149~MeV. At this energy the propagation depth exceeds by far the dimension of the sample under study: from the simulation results in Fig.~\ref{fig:Simulation}c, the number of created vacancies, and consequently NV centers, remains high after a propagation of more than 10~cm inside the diamond. 
Multiplying the profile by the beam fluence gives the profile of created vacancy density, the knowledge of which helps to set the irradiation parameters for a desired application. 

\begin{figure}[htb!]
    \centering
    \includegraphics[width=1\textwidth]{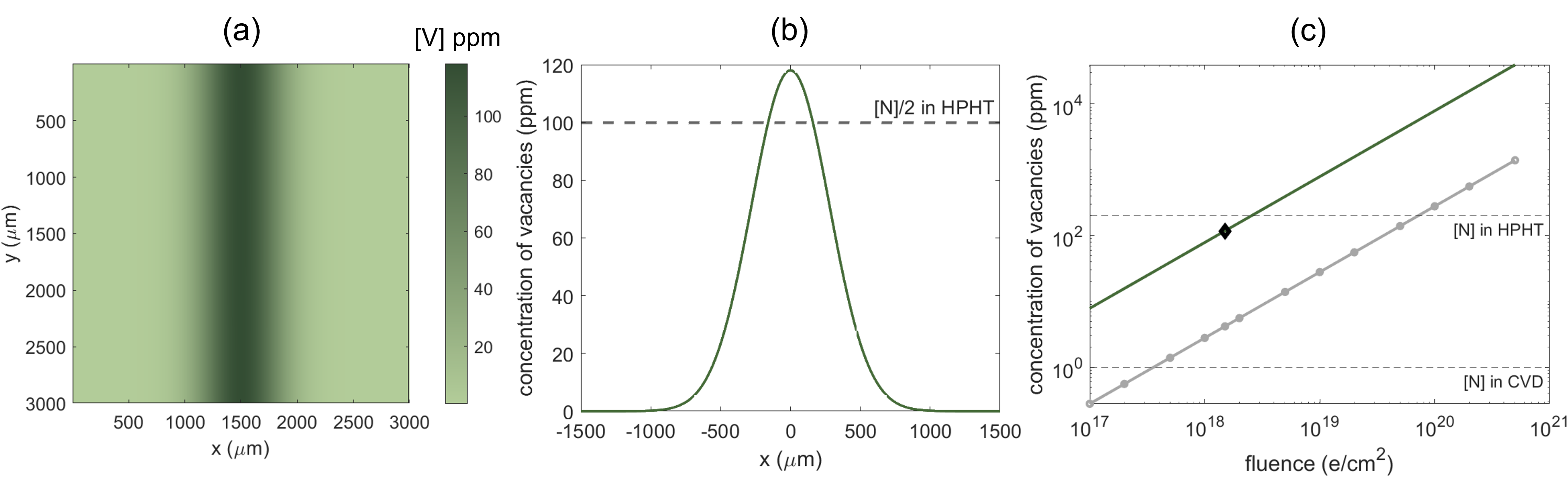}
    \caption{a-b) Simulation of the vacancy concentration (in ppm) generated by 155 MeV electrons (500~$\mu$m gaussian beam, maximum fluence $\sim$ 1.5$\cdot 10^{18}~\mathrm{e}^-/\mathrm{cm}^2$). a) xy-map, for the beam propagating in the y direction and b) lateral profile, compared with half the [N] present in the substrate.  c) Vacancy concentration (in ppm) generated at different irradiation fluences by 200~keV electrons (gray line) and 155~MeV electrons (dark green line). The horizontal lines correspond to half the [N] present in HPHT and CVD substrates, respectively.} 
    \label{fig:VacanciesConcentration}
\end{figure}

For a fluence of 1.5$\cdot 10^{18}~\mathrm{e/cm}^2$, the concentration of vacancies generated by a 155~MeV gaussian beam (diameter $\sim$500~$\mu$m) propagating for 3~mm in a diamond sample is shown in Fig.~\ref{fig:VacanciesConcentration}a. We predict no significant attenuation along the y-axis across the 3~mm path. The lateral profile (Fig.~\ref{fig:VacanciesConcentration}b) shows that for the chosen fluence the vacancy concentration is of similar magnitude as the nitrogen concentration in typical HPHT substrates, estimated by commercial providers as  [N]$\sim 200$~ppm. The gaussian beam profile allows to explore the impact of different fluences into the same diamond substrate. 
In comparison, Fig.~\ref{fig:VacanciesConcentration}c shows the concentration of vacancies obtained for a wide range of fluences. The gray line refers to the low energy irradiation regime (200~keV), while the 155~MeV case is reported in black. The horizontal dashed lines represent the [N] in the two considered substrates  (CVD and HPHT respectively).

\section{Irradiation experiments}

\subsection{Extremely high energy electron irradiation}
\begin{figure}
    \centering
    \includegraphics[width=0.9\textwidth]{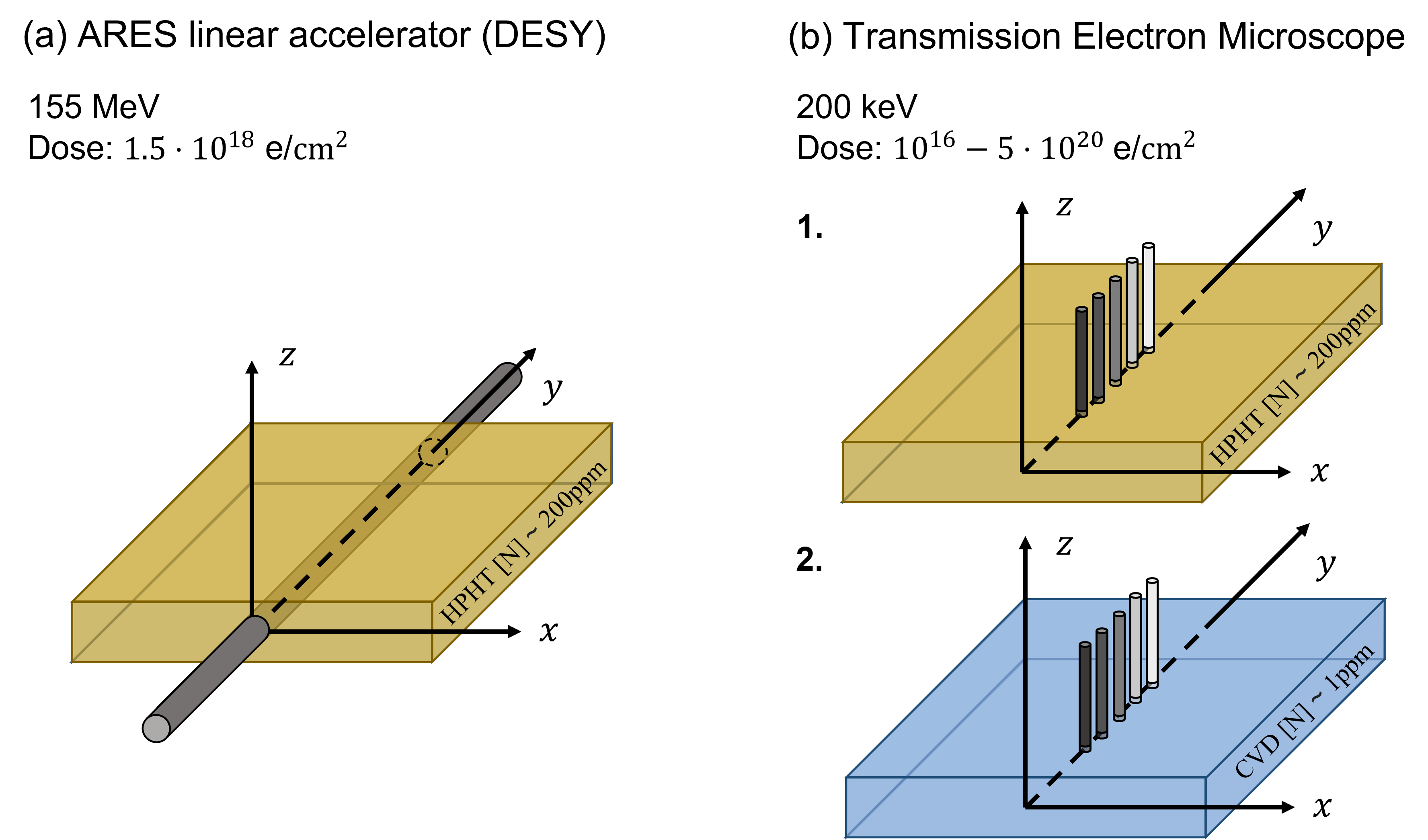}
    \caption{Schematic of the electron irradiation experiments using a) 155~MeV electrons in a $\sim$500~$\mu$m beam with a fluence of $1.5 \cdot 10^{18}$~e/cm$^2$ at ARES (DESY) and b) 200~keV electrons in a 15~$\mu$m beam using a TEM. Ten different locations with fluences in the range of $1 \cdot 10^{16}$ to $5 \cdot 10^{20}$~e/cm$^2$ were irradiated on each of two samples (1. HPHT and 2. CVD), presenting a different initial nitrogen concentration.}\label{fig:IrradiationSchematic}
\end{figure}

The 155~MeV electron irradiation took place at ARES (Accelerator Research Experiment at SINBAD) at DESY, Hamburg. ARES is a normal-conducting linear accelerator for research and development purposes with ultra-short electron bunches. This experiment was installed at the in-air experimental station located at the end of the accelerator. 
A dedicated holder was designed in order to properly align the sample to the beam, with 1~$\mu$m accuracy.  The experiment was performed on a commercial HPHT diamond sample from Element6 (3$\times$3$\times$0.3~mm$^3$, [N]$\sim$~200 ppm). 
The beam presents a gaussian profile and passes through the sample along the $y-$axis, as schematically reported in Fig.~\ref{fig:IrradiationSchematic}a. The diameter of the beam is estimated to be $2r \sim (500 \pm 50)$~$\mu$m. 

According to the results presented in Fig.~\ref{fig:VacanciesConcentration}a and discussed in the previous section, we set the target fluence to $1.5 \cdot 10^{18}$~$\mathrm{e/cm}^2$. To reach this value it was crucial to operate at very high bunch charges, up to 120~pC per pulse, with a repetition rate of 10~Hz. The sample was irradiated for 96 hours and the beam position was monitored regularly.
Following the irradiation, a conventional low pressure thermal annealing step (800$^\circ$C during 4~h under P=$10^{-6}$~mbar) was performed. 
Both the annealing and the following measurements were carried out at the École Polytechnique Fédérale de Lausanne (EPFL).
In the following, this sample is referred to as Sample 1.

\subsection{Low energy electron irradiation}
In parallel, we performed a more standard irradiation process, using low energy electrons from a TEM available at the Interdisciplinary Centre for Electron Microscopy - EPFL. We used a Talos F200S TEM providing electrons at 200~keV within a homogeneous beam of 15~$\mu$m diameter. 
The irradiation geometry is reported in Fig.~\ref{fig:IrradiationSchematic}b. Moving the sample while changing the beam current and the irradiation time allowed us to easily control the irradiation fluence from one region to the other, within the same sample. As reported in Fig.~\ref{fig:VacanciesConcentration}c, tested fluences cover a wide range from $1 \cdot 10^{16}$ to $5 \cdot 10^{20}$~e/cm$^2$. The irradiation time for the highest fluence was around 15 minutes, with a flux density up to $1 \cdot 10^{18}$~e/s/cm$^2$. 
Two different samples were subject to the same irradiation procedure: an HPHT sample from Element6 (3$\times$3$\times$0.3~mm$^3$, [N]$\lesssim$200~ppm, belonging to the same batch as Sample 1), and a CVD sample from Appsilon (1$\times$3$\times$0.3~mm$^3$, [N]$\lesssim$1~ppm). Also in this case, following  irradiation, the same conventional low pressure thermal annealing step as above was performed. 

\begin{figure}
    \centering
    \includegraphics[width=0.7\textwidth]{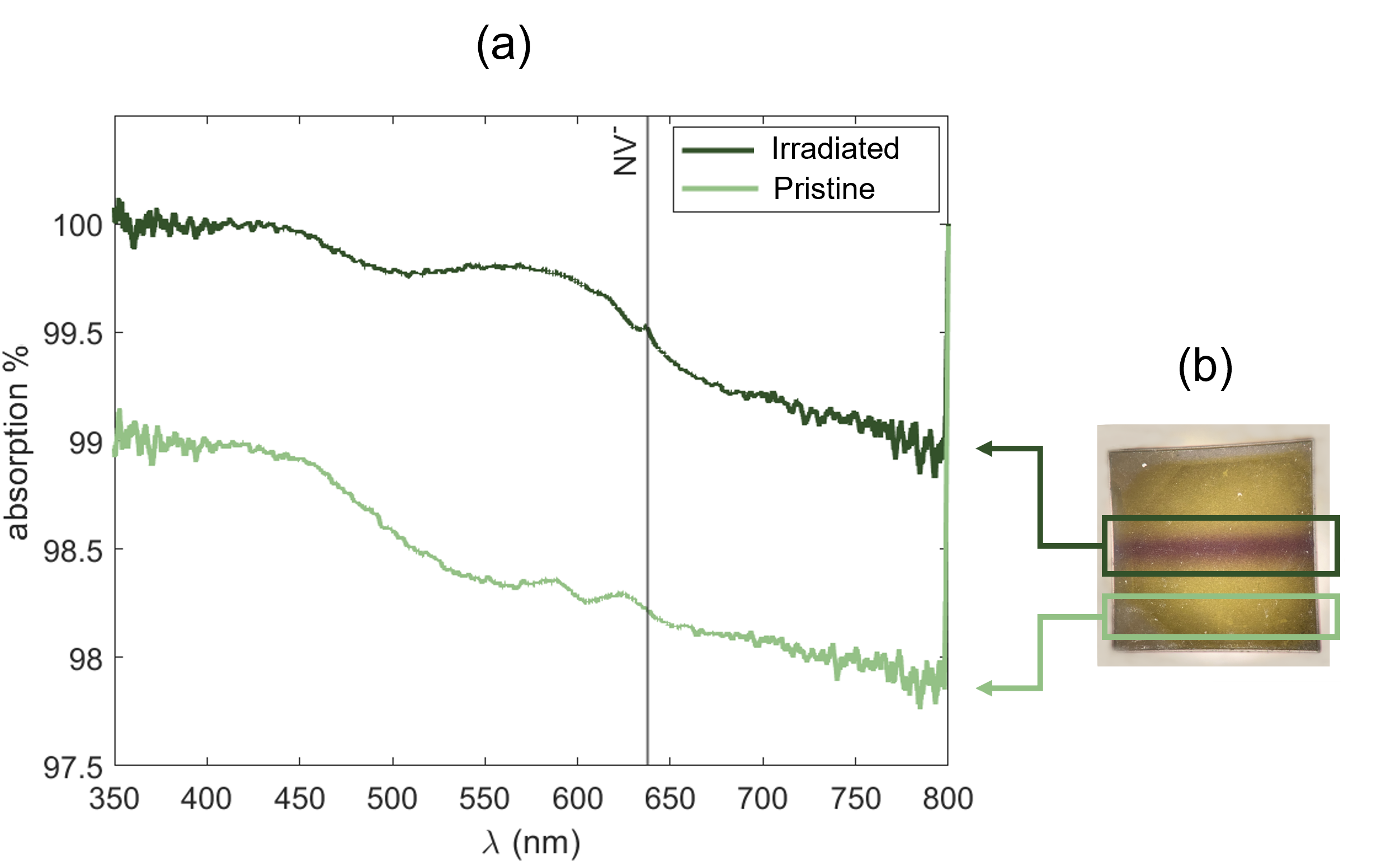}
    \caption{ a) UV/Visible absorption spectroscopy. The light green curve corresponds to the pristine region, the dark green curve to the irradiated one. Data are acquired from 0.2x3~mm$^2$ stripes. b) Optical picture of the sample after irradiation: a dark line corresponding to the irradiated region is clearly visible.}\label{fig:Absorption}
\end{figure}

\section{Results and discussion}

The simulation we develop describes well the penetration depth and gives us an estimate of the concentration of created vacancies, thus offering a guide in setting the irradiation parameters and an important tool to interpret the results.
In the following we carefully characterize, with different techniques, the obtained samples, in order to experimentally explore the impact of electron irradiation on the crystal lattice and investigate the properties of the created NV$^-$, having the quantum sensing applications in mind. 

\subsection{Extremely high energy electron irradiation}

Important changes in the diamond sample induced by the irradiation and annealing processes can already be appreciated by looking at the sample with naked eyes, as reported in Fig.~\ref{fig:Absorption}b. In particular, the central region, corresponding to the highest irradiation fluence, $1.5 \cdot 10^{18}$~e/cm$^2$, presents a darker color than the non-irradiated region at the edge of the specimen. We refer to the latter as the pristine region.  
To better investigate the nature of the defects created during irradiation and annealing, we first compare the UV/Visible absorption spectrum at the position of the highest fluence with the one from the pristine region. The results are reported in Fig.~\ref{fig:Absorption}a. The light green line corresponds to the pristine region and presents a yellow transmission window, typical in HPHT diamonds due to the high N concentration and responsible for their yellow color \cite{green2022diamond}. 
The dark green curve was acquired in the highest fluence region: the NV$^-$ absorption peak at 638 nm and its phonon sideband appear, clearly showing that the process induces an enhancement of the NV$^-$ concentration.
This spectrum can be decomposed in the NV$^-$ and its sideband plus a broad absorption spectrum, responsible for the dark color of the irradiated region. This second feature could be associated with large vacancy clusters, which may be produced by the ultra-high energy electron irradiation process \cite{green2022diamond,fujita2009large}.
Note that the spatial resolution of these UV/Visible absorption measurement is poor: each curve in Fig.~\ref{fig:Absorption} is an average over a 0.2x3~mm$^2$ stripe.

\begin{figure}
    \centering
    \includegraphics[width=1\textwidth]{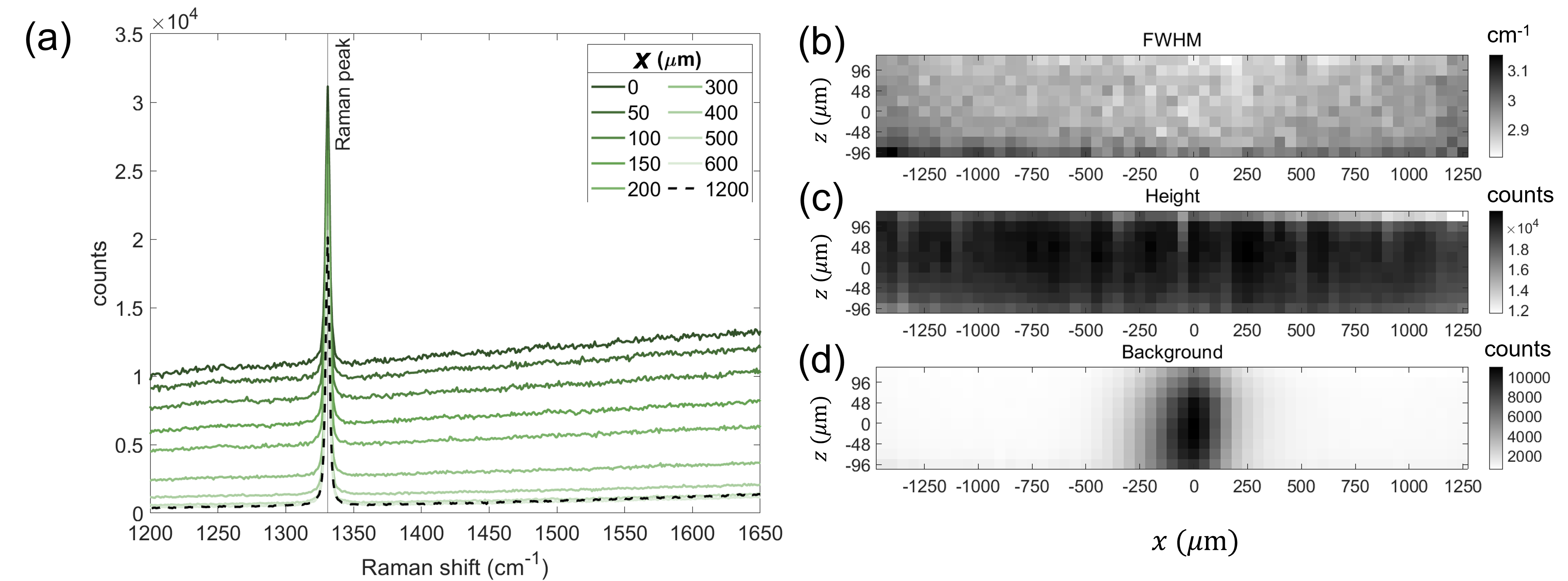}
    \caption{Raman spectroscopy under 785~nm excitation wavelength. a) Raman spectra for different positions $x$. Lorentzian fits of the main peak are used to extract $xz$-map of its FWHM (b),  amplitude (c) and vertical offset (d).}\label{fig:Raman785}
\end{figure}

To gain more insight into possible structural damages of the crystal lattice, we use Raman spectroscopy~\cite{knight1989characterization,prawer2004raman}. 
The results obtained under 785~nm laser excitation are reported in Fig.~\ref{fig:Raman785}. The Raman spectra presented in Fig.~\ref{fig:Raman785}a are acquired at different $x$, for $z \sim 0$ and $y \sim 1500~\mu$m (refer to Fig.~\ref{fig:IrradiationSchematic}a for the coordinate system). The dashed line corresponds to the spectrum acquired in a pristine region of the sample. 
For each spectrum, the main peak is fitted with a Lorentzian. Variations of the fit parameters (full width half maximum (FWHM), height and background offset) within a $xz-$plane transverse to the irradiation beam (at $y \sim 1500~\mu$m) are reported in Figs.~\ref{fig:Raman785}b-d. We notice that the higher the irradiation fluence the higher the background, while the height and width do not significantly change. The FWHM agrees with typical values measured in pristine diamond, and no Raman peak downshift is observed, nor presence of "damage" peaks (typically reported at 1490 and 1630 cm$^{-1}$). We did not observe any dependence of these features over the longitudinal $y-$ axis. These results indicate that the treatment did not heavily affect the crystal structure \cite{orwa2000raman}. The high background in the irradiated region could be due to the creation of defects exhibiting PL in the infrared. Vacancy-related defects such as large clusters of vacancies may be involved.

\begin{figure}
    \centering
    \includegraphics[width=0.6\textwidth]{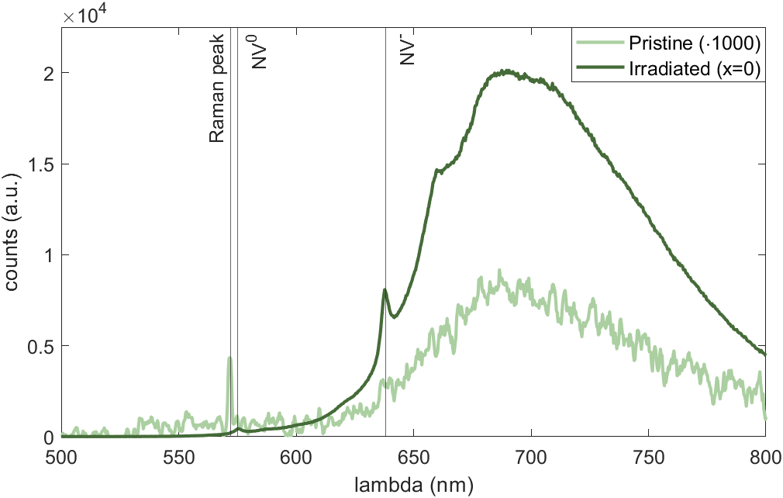}
    \caption{Photoluminescence spectroscopy under 532~nm excitation wavelength. PL spectra from the maximally irradiated area (dark green curve) and the pristine one (multiplied by 1000, light green curve) are compared.}\label{fig:PL1}
\end{figure}

We now investigate the properties of the created NV centers using photoluminescence (PL) spectroscopy and optically detected magnetic resonance (ODMR). The two charge states, NV$^-$ and NV$^0$, present different PL spectra with a characteristic zero phonon line (ZPL) at 575~nm and 638~nm, respectively. Relying on a calibrated sample, PL spectroscopy can also be used for a quantitative estimate of the concentrations of the different defects. 
In Fig.~\ref{fig:PL1}, PL spectra obtained under 532~nm excitation wavelength in the maximally irradiated region and in the pristine one are compared. Note that the spectrum from the pristine region is multiplied by 1000. 

In both spectra, we clearly recognize the contribution from NV$^-$ centers, with the ZPL peak at 638~nm and associated broad phonon sideband peaking around 700~nm.
A smaller peak, at 575~nm, is also visible in the spectrum of the irradiated region and corresponds to the ZPL from NV$^0$ centers. 
The mean PL increases by several orders of magnitude after the irradiation plus annealing process, indicating successful creation of NV$^-$ with high density. More precisely, comparing the spectra at different stages of the process (before irradiation, after irradiation and after irradiation plus annealing, see Fig.~\ref{fig:PL11} in the Supplementary Material), we confirm that, even though the irradiation process already induces change in the PL, the annealing step is crucial to drastically increase the number of NV$^-$ into the sample. This is explained by the mobility of the vacancies at high temperature, able to migrate over several nanometers and couple with the substitutional nitrogen atoms to form NV centers.

Fitting the NV$^-$ ZPL peak after background subtraction and comparing the peak area with the one from an independently calibrated sample, the absolute NV$^-$ concentration can be estimated \cite{acosta2009diamonds}. 

\begin{figure}
    \centering
    \includegraphics[width=1\textwidth]{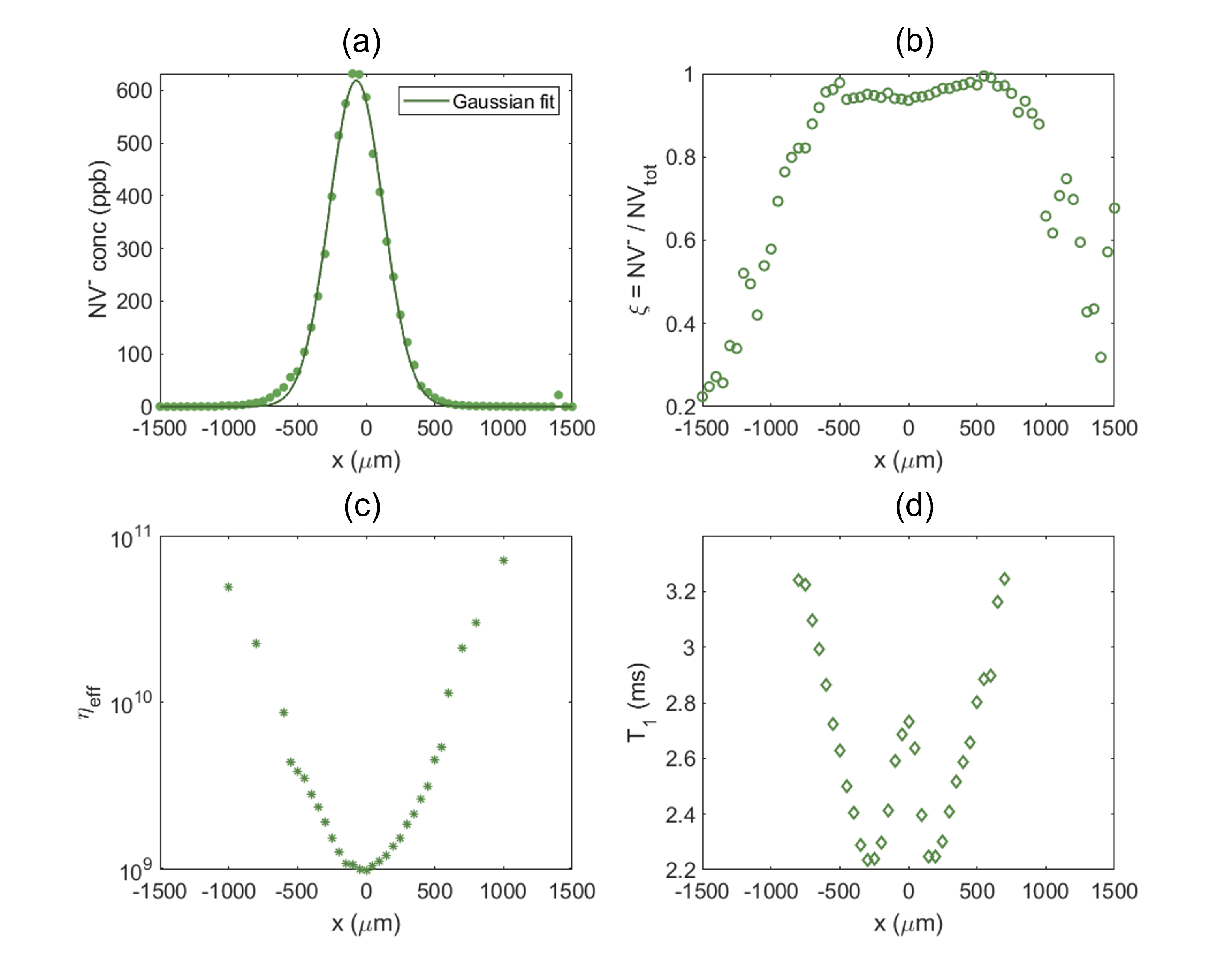}
    \caption{Plot of relevant quantities versus $x$, the distance from the maximal irradiation region. a) NV$^-$ concentration, fitted with a gaussian curve. b) Charge state conversion efficiency, $\xi$, according to Eq.~\ref{xiDW}. c) Effective shot-noise sensitivity, according to Eq.~\ref{eta}. d) Values of the longitudinal spin relaxation time T$_1$. 
    }\label{fig:All}
\end{figure}

The [NV$^-$] at varying $x$ is reported in Fig.~\ref{fig:All}a. It reaches a maximum of 0.6~ppm at $x=0$. 
A gaussian curve fits well the measured density profile, with a FHWM of $462 \pm 10~\mu$m. 
The same measurement was repeated at different positions on the $y-$axis, without significant differences, confirming that the effects of irradiation are constant over the electron beam propagation axis inside the diamond. In particular, no beam broadening was observed and the generation of NV centers appears constant at different depths. This is in good agreement with our model (Fig.~\ref{fig:VacanciesConcentration}a), which shows an almost negligible decrease of generation of vacancy density from 118~ppm to 117~ppm between the entrance and the exit of the high energy beam in the diamond crystal.

For NV sensing applications, a high concentration of NV$^-$ is not enough to achieve high performance. The presence of the neutral form of the defect, NV$^0$, is detrimental for sensing as it causes background luminescence thus reducing the ODMR contrast \cite{barry2020sensitivity}. Therefore a high [NV$^-$] / [NV$_{tot}$] ratio is desirable. We refer to this figure of merit as charge state conversion efficiency $\xi$ and compute it using the Debye-Waller decomposition method proposed in \cite{alsid2019photoluminescence}, which neglect the presence of NV$^{+}$ and takes into account the different response of NV$^{-}$ and NV$^{0}$ to 532~nm light:
\begin{equation}\label{xiDW}
    \xi=\frac{I_{ZPL,-} e^{S_-}}{k_{532}I_{ZPL,0} e^{S_0}+I_{ZPL,-} e^{S_-}}
\end{equation}
where $I_{ZPL,-}$ and $I_{ZPL,0}$ are the zero-phonon-line areas for NV$^-$and NV$^0$, respectively, weighted by their Debye-Waller factors $S_-=4.3$ and $S_0=3.3$. $k_{532}=2.5$ takes into account the different PL rate of the two charge states under 532~nm illumination.
The resulting $\xi$ is reported in Fig.~\ref{fig:All}b: in the irradiated region more than 90\% of the NV centers are in the negatively charged state.
The maximum value of $\xi$ is at $x \sim 500$~$\mu$m. The slightly decreased charge state conversion efficiency around $x \sim 0$~$\mu$m may indicates that, at this location, there are not enough electron donors to efficiently convert NV$^0$ into NV$^-$. As suggested by the simulation shown in Fig.~\ref{fig:VacanciesConcentration}b, it could be due to a concentration of created vacancies too high for the available nitrogen content. Moreover, the presence of more defects behaving as electron traps is also possible.

Many NV-sensing protocols are based on the acquisition of ODMR spectra \cite{rondin2014magnetometry,schirhagl2014nitrogen,segawa2022nanoscale,bernardi2017nanoscale} and T1 relaxometry \cite{tetienne2013spin, simpson2017electron, barton2020nanoscale, sigaeva2022diamond}. The performance of these protocols are related not only to the NV$^-$ density but also to their spin lifetimes and coherence properties, which are affected by the specific crystal lattice environment. 
We acquired continuous wave ODMR spectra in our home-made confocal microscope. More details on the set-up can be found in \cite{babashah2022optically}. We used a permanent magnet to lift the zero-field splitting degeneracy and address a single transition.

\begin{figure}
    \centering
    \includegraphics[width=0.5\textwidth]{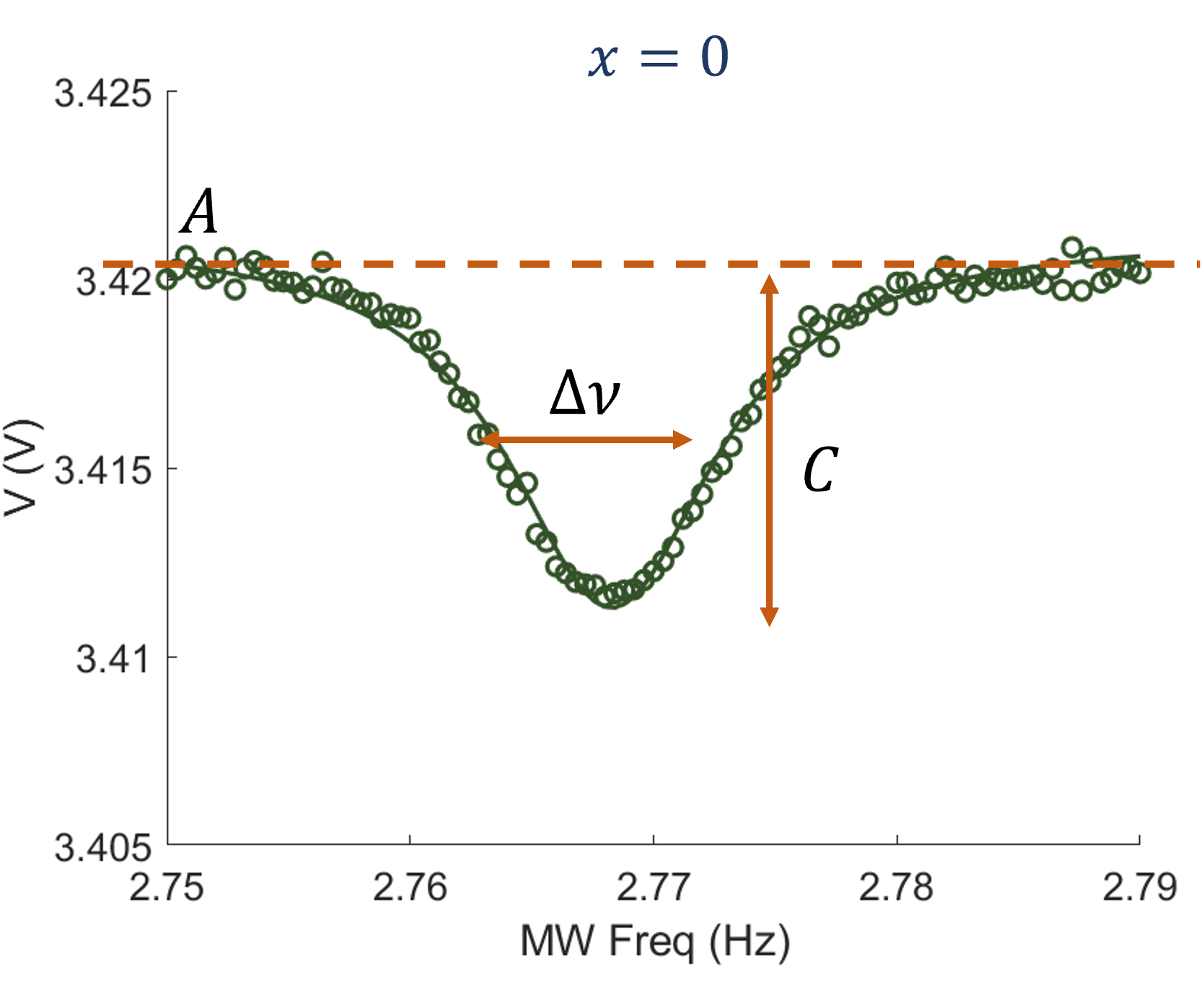}
    \caption{Optically detected magnetic resonance (ODMR) spectrum, fitted with a Lorentzian curve, from the maximally irradiated region ($x=0$). To remove the degeneracy a bias magnetic field B $\sim$ 5~mT was applied.}\label{fig:ODMRfit}
\end{figure}

An example of ODMR spectrum measured at the maximally irradiated region, $x=0$, is reported in Fig.~\ref{fig:ODMRfit}.  
From the ODMR spectrum it is possible to estimate the shot-noise sensitivity to magnetic field according to the formula \cite{dreau2011avoiding}: 
\begin{equation}
    \eta \mathrm{(T/}\sqrt{\mathrm{Hz}}) = \frac{4}{3\sqrt{3}} \frac{h}{g \mu_B} \frac{\delta \nu }{C \cdot \sqrt{R}},
\end{equation}
where $\delta \nu$ is the Lorentzian FWHM, $C$ its contrast and $R$ the rate of detected photons. The numeric prefactor depends on the resonance lineshape, which is assumed here to be a Lorentzian. $\frac{h}{g \mu_B} = 2.80 ~ \mathrm{MHz/G}$ is the NV gyromagnetic ratio.
To simplify the discussion, we introduce an effective sensitivity that does not take into account the collection efficiency of the setup:
\begin{equation}\label{eta}
    \eta_{eff} \mathrm{(a.u./}\sqrt{\mathrm{Hz}}) = \frac{\delta \nu }{C \cdot \sqrt{A}} \propto \eta \mathrm{(T/}\sqrt{\mathrm{Hz}}),
\end{equation}
where both $\delta \nu$ and $C$ are extracted fitting the ODMR spectrum with a Lorentzian (see Fig.~\ref{fig:ODMRfit}). $A$ is the voltage measured by the photodiode, proportional to the PL count rate. In the Supplementary Material (see Fig.~\ref{fig:ODMRfitPar}b,c and d), the linewidth, contrast and amplitude extracted from the fit for different positions $x$ are reported. No significant trend is visible in the linewidth, some modulation is present for the contrast that follows well the behavior of the charge state ratio $\xi$, while the PL presents a clear increase around $x=0$~$\mu$m as expected.
The effective sensitivity  $\eta_{eff}$ at different distances from the highest irradiation line is reported in Fig.~\ref{fig:All}c. We demonstrate an improvement of $\sim$40 times compared to the pristine region. The sensitivity enhancement is mainly dominated by the PL enhancement, while the variations in contrast and linewidth have a lesser impact.

The same set-up is used to estimate T$_1$ across the sample. We use the standard all-optical measurement protocol, reported for example in \cite{babashah2022optically}: a first laser pulse is used to polarize the NV$^-$, a second pulse allows to read out the spin state after a variable time delay $\tau$. T$_1$ is obtained by fitting the decay of readout PL versus $\tau$ with a single exponential function.
In Fig.~\ref{fig:All}d, the extracted T$_1$ values at different distances from the maximally irradiated region are reported.
Interestingly, T$_1$ presents a non-monotonous behaviour with two minima at $x \sim \pm 250$~$\mu$m and a local maximum at $x \sim 0$~$\mu$m. This is in contrast with the findings of \cite{jarmola2015longitudinal}, where a monotonous reduction of T$_1$ was observed when increasing irradiation fluence. However, the study from \cite{jarmola2015longitudinal} was based on CVD diamond irradiated with low energy 200~keV electrons. 
We hypothesize that other defects, such as vacancy clusters, introduced under 155~MeV irradiation may be responsible for the anomalous evolution of T1 in the highly irradiated region. However, the mechanism behind the rise of T1 remains to be clarified in future studies.

\subsection{Low energy electron irradiation}

\begin{figure}
    \centering
    \includegraphics[width=1\textwidth]{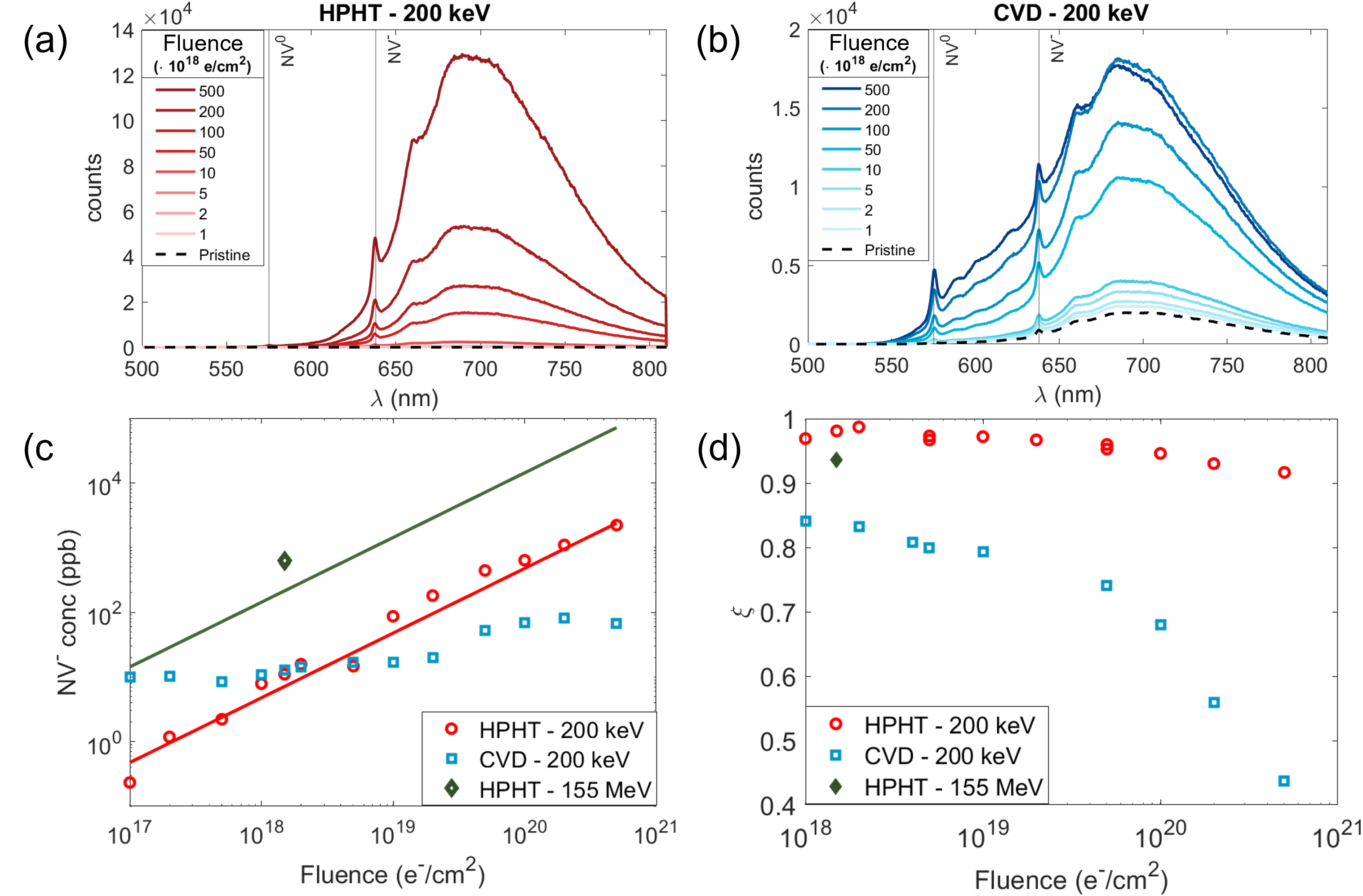}
    \caption{ a-b) PL spectra for different irradiation fluences under 200~keV electrons from the HPHT (a) and the CVD samples (b). Note the different vertical scales. c) NV$^-$ concentration as a function of irradiation fluence for the HPHT (red circles) and the CVD (blue squares) diamonds. Experimental data are linearly fitted (red line), assuming same slope as the one predicted by the simulation (Fig.~\ref{fig:VacanciesConcentration}c). The dark green line corresponds to the theoretical values for the 155~MeV case, assuming the same [V] to [NV$^-$] conversion yield as in the 200~keV case. d) $\xi$ as a function of irradiation fluence for the HPHT (red circles) and the CVD (blue squares) diamonds. The experimental values corresponding to Sample 1 are reported as a diamond symbol in panels (c,d).}\label{fig:HPHTvsCVD}
\end{figure}

We now turn to the discussion of the two samples irradiated under TEM by 200~keV electrons. 
In this case, UV/Visible absorption spectroscopy is not possible due to its poor spatial resolution vs. the small dimension of the irradiated areas. We use PL spectroscopy to estimate the NV concentration and the charge state conversion efficiency $\xi$ in the different irradiation areas. Fluences in the range $1 \cdot 10^{16}$ to $5 \cdot 10^{20}$~e/cm$^2$ are investigated. However, the irradiated areas could only be clearly identified by their PL signal for fluences higher than $1 \cdot 10^{17}$~e/cm$^2$ (resp. $1 \cdot 10^{18}$~e/cm$^2$) for the HPHT (resp. CVD) sample. 

In Fig.~\ref{fig:HPHTvsCVD}a-b, we report the spectra corresponding to different irradiation fluences, for both the HPHT and CVD substrates. The Debye-Waller decomposition analysis presented in the previous section allows us to quantify the NV$^-$ concentration and charge state ratio $\xi$ as a function of irradiation fluence for both samples (Fig.~\ref{fig:HPHTvsCVD}c-d).
In the HPHT sample, where the initial nitrogen concentration is higher, increasing the fluence leads to a continuous and approximately linear increase in PL intensity, i.e. in NV$^-$ concentration, with a maximum PL enhancement of $\sim$ 4 order of magnitude and a maximum [NV$^{-}$]$\sim$ 3~ppm, without any sign of saturation for the explored range of fluences (red circles in Fig.~\ref{fig:HPHTvsCVD}c). Experimental data are linearly fitted, assuming same slope as the theoretical curves reported in Fig.~\ref{fig:VacanciesConcentration}c. 
Comparing the measured [NV$^-$] with the simulated [V], we infer that the conversion yield from created vacancy to NV$^-$ is $\sim 1.7 \cdot 10^{-3}$.
The excess of nitrogen in the HPHT diamond provides sufficient electron donors to ensure a high [NV$^-$]/[NV$_{tot}$] ratio ($\xi>0.9$, red circles in Fig.~\ref{fig:HPHTvsCVD}d).

On the contrary, for the CVD sample, we observe a sublinear increase of PL intensity vs. irradiation fluence with a saturation at the highest fluence values (blue squares in Fig.~\ref{fig:HPHTvsCVD}c).
Moreover, a growing contribution from the NV$^0$ peak to the PL spectrum reflects a decrease in the charge state conversion efficiency $\xi$ for 0.85 to 0.4 (blue squares in Fig.~\ref{fig:HPHTvsCVD}d).
These observations confirm that the nitrogen content is a limiting factor for increasing [NV$^-$] following electron irradiation and annealing, as it limits the charge state conversion efficiency.
 
From Fig.~\ref{fig:Simulation}c-d, the simulated number of created vacancies, close to the surface, at 155~MeV exceeds the number of created vacancies at 200~keV by a factor of roughly 30 (dark green vs red lines in Fig.~\ref{fig:HPHTvsCVD}c). Experimentally, we measure a difference in [NV$^-$] of $\sim$ 60 times for the same HPHT diamond and same irradiation fluence (black diamond in Fig.~\ref{fig:HPHTvsCVD}c), which is in qualitative agreement with the simulation. The factor 2 of discrepancy can be explained by uncertainty in the irradiation fluence and/or different vacancies formation and recombination mechanisms in the two energy regimes. 
We notice a slightly lower value of $\xi$ for the 155~MeV process with respect to 200~keV on the HPHT sample (black diamond in Fig.~\ref{fig:HPHTvsCVD}d). It is possible that the ultra high energy irradiation creates other vacancy-containing defects that also behave as electron acceptors, thus lowering the fraction of NV$^-$. As shown in Fig.~\ref{fig:All}b, moving few micrometers apart from the maximally irradiated region, $\xi$ increases, reaching values higher than 0.98. The current simulation provides the concentration of vacancies, but cannot infer anything about their possible aggregation states. This aspect may be relevant and should be addressed in a dedicated work.

\section{Conclusion and outlook}

In this work, we have explored the impact of ultra high energy electron irradiation (155~MeV) on diamond with a focus of its application for NV-based quantum sensing. We developed a simulation model able to faithfully describe this regime, we performed irradiation and annealing of an HPHT diamond substrate with large nitrogen content, characterized the NV$^-$ properties in the obtained sample, and compared the results with those obtained under low energy (200~keV) electron irradiation.
The main finding is that 155~MeV irradiation yields approximately 60 times higher NV$^-$ concentration compared to 200~keV irradiation of the same diamond at the same fluence, while maintaining near ideal charge state conversion ratio ($\xi>0.95$).
Even though an increase in broadband absorption and some background emission was observed, possibly due to an increased density of vacancy complexes, their effect on estimated ODMR sensitivity was not significant. 
The main advantage of the extremely high energy electrons is their huge penetration depth, estimated here to be more than ten centimeters. Accordingly, this approach may be used in the future for batch irradiation of hundreds of substrates stacked together, with a potential for cost reduction (provided that an affordable source of ultra high energy electrons is accessible). 
This energy range may also present some advantages when irradiating diamond-containing heterostructures in order to reduce electron absorption and/or scattering in other layers. 

Due to the limited access to the DESY irradiation facility we could explore only one irradiation energy and fluence. It would be interesting to vary these parameters, in order to better understand their impact and optimize the NV$^-$ creation while reducing the induced lattice damage. Our experiments and simulations also show that [NV$^-$] after irradiation and annealing is limited by [N] in the initial substrate, with a maximal conversion efficiency from nitrogen to NV$^-$ limited to 0.3\%. This could be improved with in-situ annealing, as suggested by previous studies on HPHT samples \cite{capelli2019increased}: with this technique a conversion efficiency higher than 1 \% should be achievable.

Finally, we observed an unexpected, non-monotonous behaviour for the $T_1$ relaxation time vs. effective irradiation fluence at 155~MeV. Further experiments and/or modelling should help understanding the nature and density of the different defects in the diamond lattice and how they may impact T$_1$. Most notably, techniques like electron-paramagnetic resonance and electron spin resonance can be valid tools for this purpose \cite{iakoubovskii2004vacancy,twitchen1999electron} and should be employed to detect the presence of vacancy-related defects, e.g. single vacancies and vacancy clusters that are likely created by the highly accelerated electrons and secondary displaced atoms.
The numerical model that we developed allows to explain our experimental observations in terms of NV density and charge state conversion efficiency, but cannot capture the formation of such vacancy-related defects. More refined models could be developed in order to better understand what is happening at the atomic level. 

\section*{Acknowledgement}
This project has received funding from the EPFL Interdisciplinary Seed Funds, from the EPFL Center for Quantum Science and Engineering, from the Swiss National Science Foundation (grants No. 98898 and 204036) and from the European Union's Horizon 2020 research and innovation program under the Marie Skłodowska-Curie grant agreement No. 754354. 
We sincerely thank the EPFL Crystal Growth Facility (especially Yoan Trolliet), for the help provided in the annealing process. 
We acknowledge Umut Yazlar from Appsilon B.V. (Delft, Netherlands) for his valuable contribution in providing single crystal CVD diamond samples with desired characteristics.

\bibliography{references}

\newpage
\section*{Supplementary Material}

\subsection*{Simulation details}

The simulation of the density of vacancies generated in diamond by an accelerated electron beam is performed using a macroscopic approach, in contrast to the Monte Carlo methods \cite{drouin2007casino,ziegler2010srim} that model individual electron trajectories and local interactions of individual accelerated particles with the carbon atoms. We implemented our model using the proprietary coding language of Gatan Microscopy Suite (GMS 3.5) software.

In a first step, the energy profile of the accelerated electron beam is calculated versus the propagation depth in diamond. The two major sources of energy losses of the incident beam are considered, namely the ionization losses and the radiative losses by bremsstrahlung. The energy losses of the fast electrons ($\Delta E$) are calculated in an iterative way, using a small (100 nm and 1 $\mu$m, respectively for the 200 keV and 155 MeV electrons) step of propagation depth ($\Delta L$) in the material, based on the definition of the stopping power. The ionization losses result from the inelastic interactions of the fast electrons with the atomic electrons of the target material. The stopping power associated to the ionization losses, $S_i$, defines the average energy loss of the incident beam per unit of propagation length. $S_i$ is given by the Bethe-Bloch formula, modified for the case of incident electrons as \cite{leo_techniques_1994}:
\begin{equation}
S_i = -\frac{\Delta E_i}{\Delta L} = 2 \pi \rho Z m_0 r_e^2 \frac{c^2}{\beta^2}.\ln\left( \frac{(\gamma -1)^2(\gamma +1)m_e^2 c^4}{2I^2} +1-\beta ^2+ \frac{\frac{\gamma ^2 \beta ^2}{8}-(2\gamma\beta +1)\ln 2}{(\gamma\beta +1)^2} -\delta -2\frac{\zeta}{Z} \right)
\label{Eq1}
\end{equation}
$\rho$ is the volume atom density of the target ($\rho = 1.76 \cdot 10^{23}$ cm$^{-3}$ for diamond), $Z$ the atomic number of the target atoms ($Z = 6$ for carbon), $m_0$ is the rest mass of an electron, $r_e$ the classical electron radius, $\beta=\frac{v}{c}$ with $c$ the speed of light in vacuum and $v$ the relativistic speed of the incident electron and $\gamma$ the Lorentz factor. $I$ is the mean ionization potential of the target atoms, 79 eV for carbon \cite{leo_techniques_1994}. $\delta$ is the density effect correction of the Bethe-Bloch equation, correcting for the influence of the incident electron on the target atom polarization and affecting primarily very high energy incident electrons. $\zeta$ is the shell correction of the Bethe-Bloch equation, affecting lower energy incident electrons when the assumption that target atomic electrons are stationary with respect to the incident particle no longer holds. The numerical expression of the marginal $\delta$ and $\zeta$ corrections is described in \cite{leo_techniques_1994}. The stopping power associated to the radiative losses by bremsstrahlung, $S_r$, is given by,
\begin{equation}
S_r = -\frac{\Delta E_r}{\Delta L} = \frac{E.\rho}{X_0}
\label{Eq2}
\end{equation}
with $E$ is the acceleration energy of the fast electrons and $X_0$ is the radiation length, expressed as \cite{leo_techniques_1994}:
\begin{equation}
X_0 = \left(716.4~g.cm^{-2}\right).\frac{A}{Z(Z+1).ln\left(\frac{287}{\sqrt{Z}}\right)}
\label{Eq3}
\end{equation}
$A$ is the mass number of the target atoms ($A = 12.01$ for carbon). $\Delta E_i$ and $\Delta E_r$ contributions to the total average energy loss per unit of propagation length are plot versus the energy of the incident electron beam in Figure \ref{fig:lossesVSenergy}, for the diamond target.  We observe that $\Delta E_i$ is predominant at low energies and $\Delta E_r$ is predominant at very high beam energies.

\renewcommand{\thefigure}{S1}
\begin{figure}[htb!]
    \centering
    \includegraphics[width=0.5\textwidth]{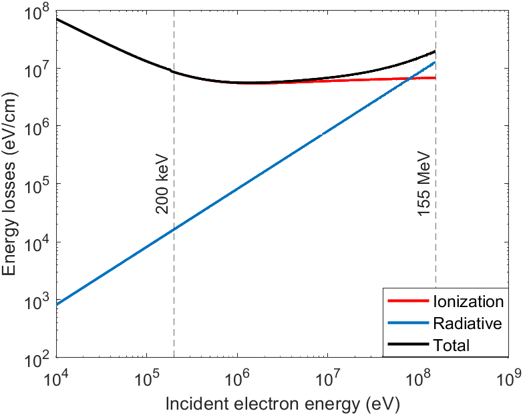}
    \caption{Contributions of the ionization (red) and radiative (blue) energy losses to the total (black) average energy losses per unit of propagation length, plot versus the energy of the incident fast electrons.}\label{fig:lossesVSenergy}
\end{figure}

\renewcommand{\thefigure}{S2}
\begin{figure}[ht]
    \centering
    \includegraphics[width=0.9\textwidth]{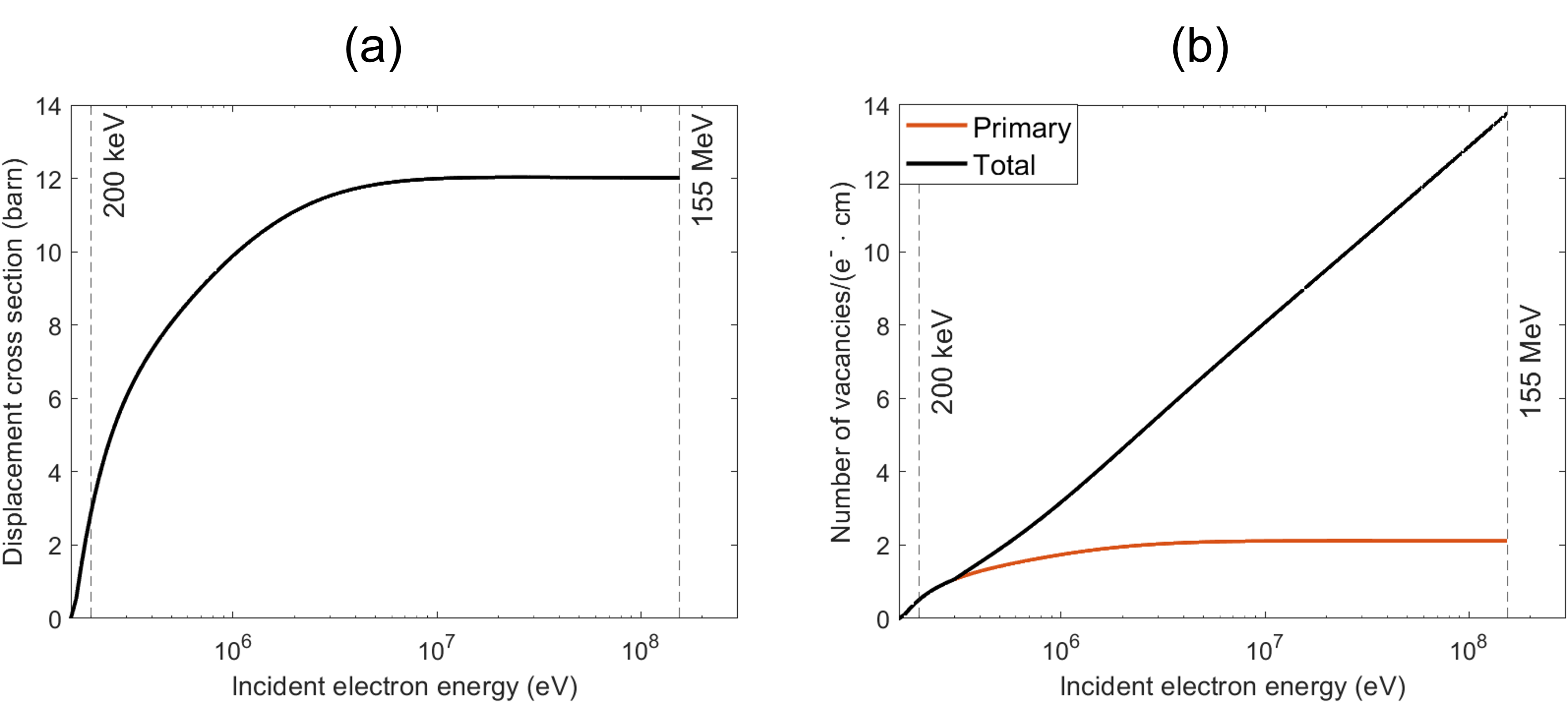}
    \caption{a) Plot of the displacement cross-section for electron irradiation of diamond versus the energy of the incident fast electrons. b) Primary (red) and final (black) number of vacancies generated in diamond by 1 fast electron versus the energy of the incident electron.}\label{fig:energyprofiles}
\end{figure}

In a second step, the number of vacancies generated by a single fast incident electron of energy E per unit of propagation length is calculated. Then the previously-calculated energy profile of the incident electron beam versus the propagation depth can be converted into a profile of vacancy density versus the propagation depth in the target material. The vacancy generation by ionization damages of the incident electron beam is neglected for N-doped diamond and only the knock-on damages resulting from ejection of an atom nuclei from the crystal lattice by energy transfer from a fast incident electron is considered. To this aim, the displacement cross-section, $\sigma_d$, is calculated. It traduces the total cross-section for the scattering of fast electrons by the atomic nuclei and generating an atom displacement. An atom is displaced from the crystal lattice, thus generating a primary vacancy-interstitial pair, if the energy transferred to the nucleus is higher than the displacement threshold energy $E_d$ (35 eV for carbon \cite{campbell2000radiation}). The total electron-nucleus scattering cross-section in the Mott theory, after integration between $E_d$ and the maximal energy transferred to the nucleus $E_{max}$, can be approximated for light elements with the McKinley and Feshback developments as \cite{mckinley_coulomb_1948, banhart_irradiation_1999},
\begin{equation}
\sigma_d = \left(0.25~barn\right).Z^2 \left(\frac{1-\beta^2}{\beta^4}\right) \left[ \frac{E_{max}}{E_d}-1+2\pi\beta\frac{Z}{137}\left(\sqrt{\frac{E_{max}}{E_d}}-1\right)-\left(\beta^2+\pi\beta\frac{Z}{137}\right).ln\left(\frac{E_{max}}{E_d}\right)\right]
\label{Eq4}
\end{equation}
where $E_{max}$ is given by \cite{egerton_electron_2011}:
\begin{equation}
E_{max} = \frac{2E(E+2m_0c^2)}{Mc^2}
\label{Eq5}
\end{equation}
$M$ is the mass of the target nucleus. The number of primary vacancy-interstitial pairs generated by a single fast incident electron per depth unit, $n$, is expressed:
\begin{equation}
n = \frac{1-e^{-\rho \sigma_d \Delta L}}{\Delta L}
\label{Eq6}
\end{equation}
When a primary recoiled target nucleus has an energy higher than $E_d$, it can further displace other nuclei from the crystal lattice, giving rise to a higher final number of stable vacancy-interstitial pairs per depth unit in the crystal, $N$, given by \cite{banhart_irradiation_1999}:
\begin{equation}
N = n\left(1+0.5\ln\frac{E_{max}}{2E_d}\right)
\label{Eq7}
\end{equation}
The factor 0.5 in front of the logarithm stands for considering that 50\% of the vacancies initially generated in diamond through cascade process recombine spontaneously, as predicted by molecular dynamics \cite{buchan_molecular_2015}. Figure \ref{fig:energyprofiles}a plot the displacement cross-section for electron irradiation in diamond versus the beam energy and shows a saturation at higher beam energies. Figure \ref{fig:energyprofiles}b plot the primary and final number of carbon vacancies generated per depth unit in diamond by a single fast electron versus the beam energy. The saturation of the generation of the primary carbon vacancies at higher beam energies is explained by the saturation of the displacement cross-section, while the total density of vacancies increases continuously by considering the cascade phenomena.

\subsection*{Supplementary figures}

\renewcommand{\thefigure}{S3}
\begin{figure}[hb]
    \centering
    \includegraphics[width=0.6\textwidth]{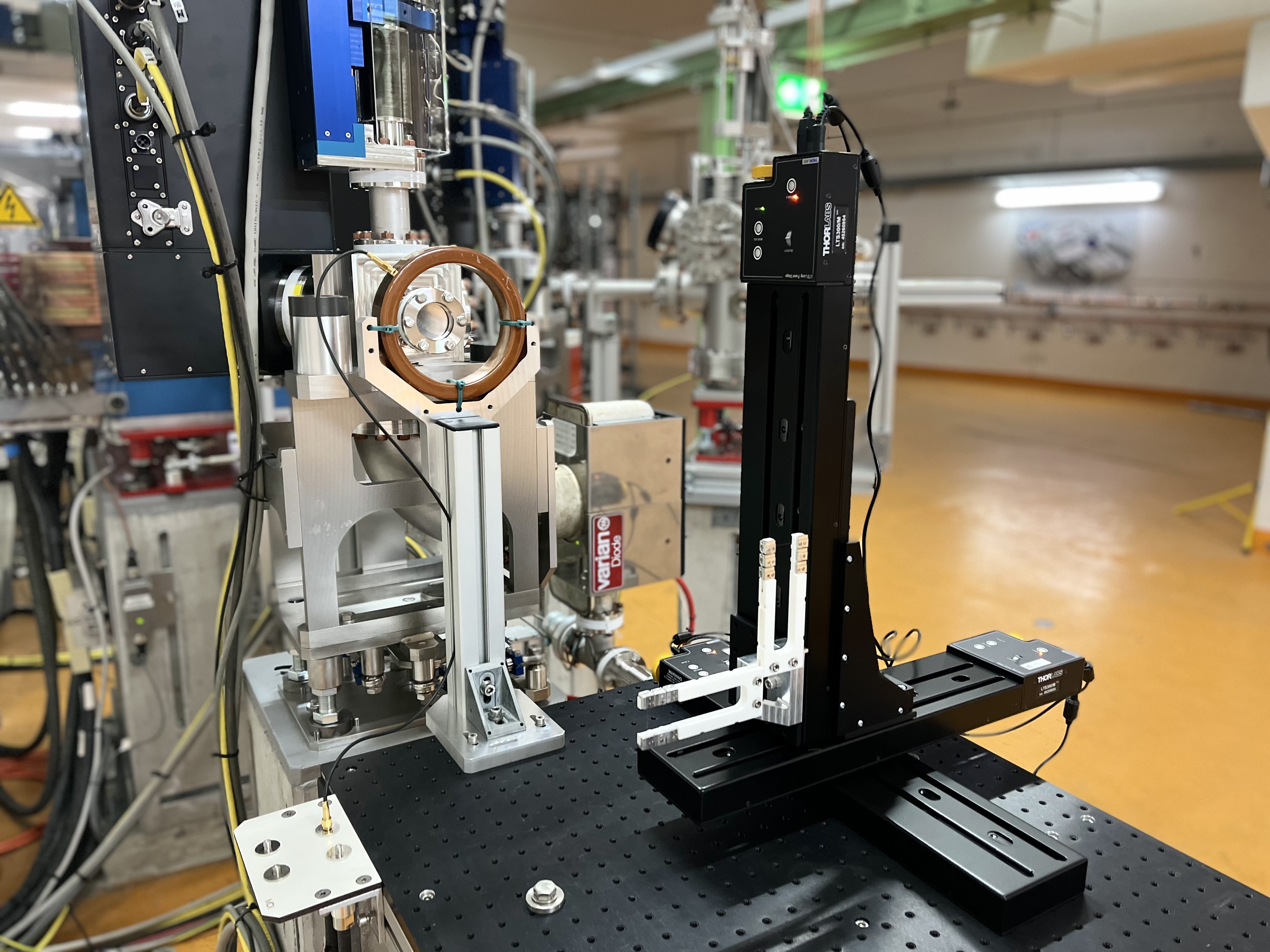}
    \caption{Picture of the irradiation point in the DESY setup.}\label{fig:DESYsetup}
\end{figure}

\renewcommand{\thefigure}{S4}
\begin{figure}[hb]
    \centering
    \includegraphics[width=0.6\textwidth]{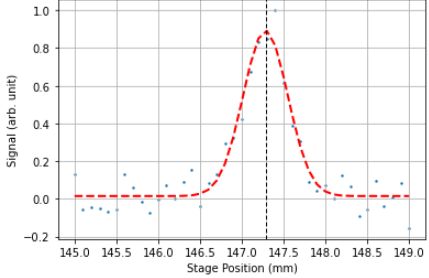}
    \caption{Beam profile - experimental data are well fitted with a gaussian profile.}\label{fig:BeamProfile}
\end{figure}

\renewcommand{\thefigure}{S5}
\begin{figure}[hb]
    \centering
    \includegraphics[width=0.6\textwidth]{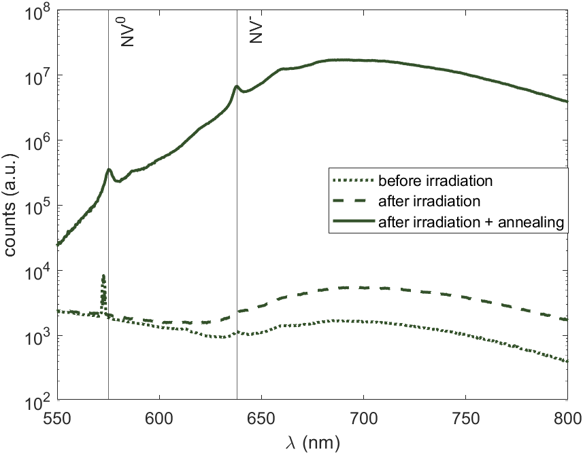}
    \caption{Photoluminescence spectroscopy for 532 nm excitation wavelength. PL spectra from the maximum irradiated area, at different moments of the process: before irradiation (dotted line), after irradiation (dashed line) and after irradiation plus annealing at 800 °C (plain line). The spectra are normalized to the Raman peak intensity.}\label{fig:PL11}
\end{figure}

\renewcommand{\thefigure}{S6}
\begin{figure}[hb]
    \centering
    \includegraphics[width=1\textwidth]{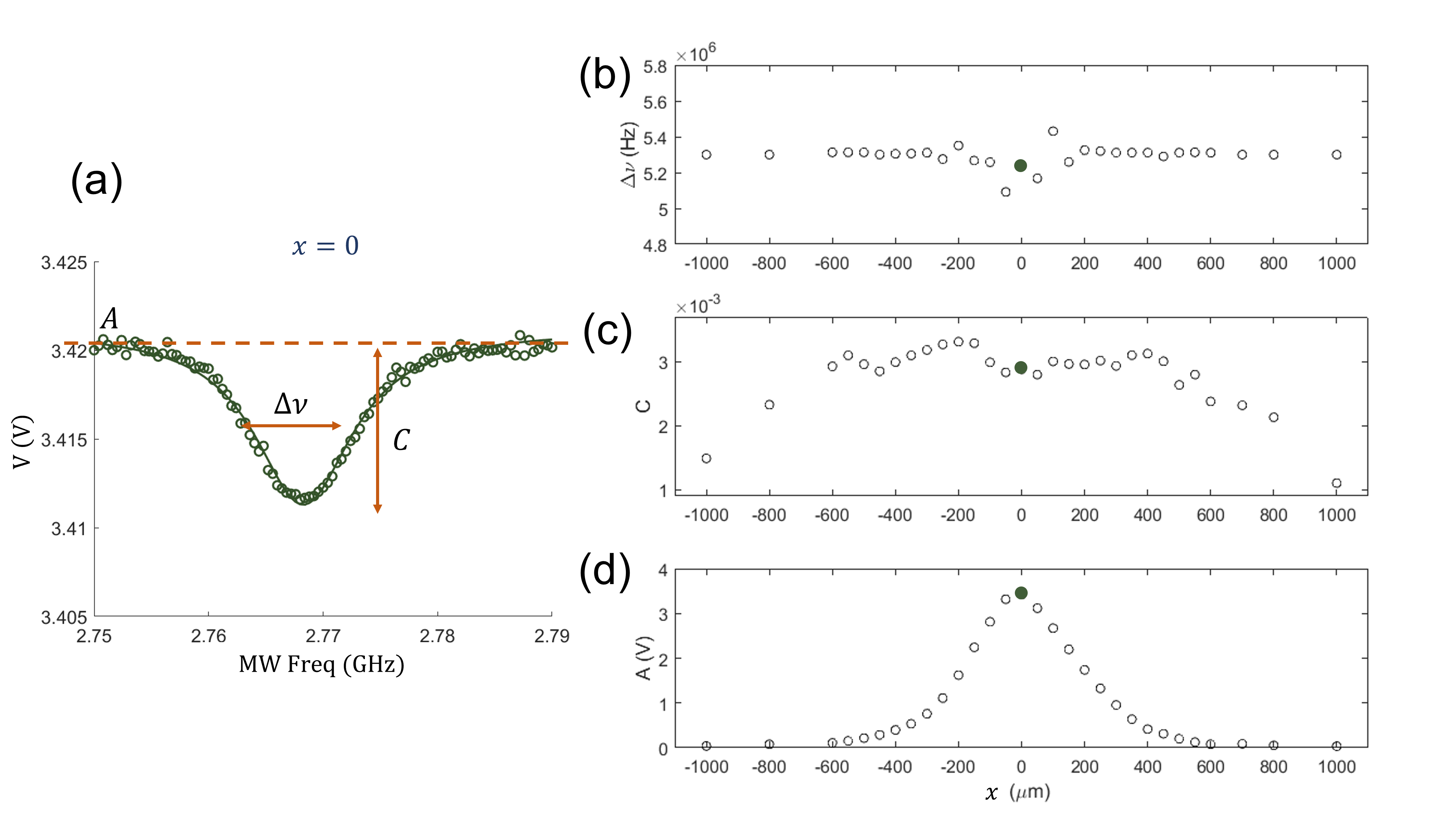}
    \caption{a) Example of ODMR spectrum and corresponding Lorentzian fit. b-c-d) Fit parameters at varying $x$.}\label{fig:ODMRfitPar}
\end{figure}

\end{document}